\documentclass[reprint,amsmath,amssymb,aps,prl,groupedaddress,nofootinbib,twocolumn,superscriptaddress]{revtex4-1}
\usepackage{graphicx}
\usepackage{amsthm,amssymb,amsmath,braket,mathdots}
\usepackage{bm}
\usepackage[pagebackref=false,pdfnewwindow=true]{hyperref} 
\usepackage{epstopdf,psfrag}
\usepackage{relsize,amsbsy}
\usepackage[export]{adjustbox}
\usepackage{makecell}
\usepackage{graphicx,xcolor,tikz}
\usepackage{float}
\usepackage{mathtools}
\usepackage[centerlast]{caption}
\usepackage{hyperref}

\newcommand{\be}{\begin{equation}}
	\newcommand{\ee}{\end{equation}}

\newcommand{\bit}{\begin{enumerate}}
	\newcommand{\eit}{\end{enumerate}}

\definecolor{bananayellow}{rgb}{1.0, 0.88, 0.21}
\definecolor{straw}{rgb}{0.32, 0.28, 0.1}

\begin{document}
	\title{Temporal disorder in spatiotemporal order}
	\author{Hongzheng Zhao}
	\affiliation{\small Max-Planck-Institut f{\"u}r Physik komplexer Systeme, N{\"o}thnitzer Stra{\ss}e 38, 01187 Dresden, Germany}
	
	\author{Johannes Knolle  }
	\affiliation{Department of Physics TQM, Technische Universit{\"a}t M{\"u}nchen, James-Franck-Stra{\ss}e 1, D-85748 Garching, Germany}
	\affiliation{Munich Center for Quantum Science and Technology (MCQST), 80799 Munich, Germany}
	\affiliation{\small Blackett Laboratory, Imperial College London, London SW7 2AZ, United Kingdom}
	
	\author{Roderich Moessner}
	\affiliation{\small Max-Planck-Institut f{\"u}r Physik komplexer Systeme, N{\"o}thnitzer Stra{\ss}e 38, 01187 Dresden, Germany}
	
	\begin{abstract}
		Time-dependent driving holds the promise of realizing  dynamical phenomenon absent in static systems. Here, we introduce a correlated random driving protocol to realize a spatiotemporal order that cannot be achieved even by periodic driving, thereby extending the discussion of time translation symmetry breaking to randomly driven systems. We find a combination of temporally disordered micro-motion with prethermal stroboscopic spatiotemporal long-range order. This spatiotemporal order remains robust against generic perturbations, with an algebraically long prethermal lifetime where the scaling exponent strongly depends on the symmetry of the perturbation, which we account for analytically.
	\end{abstract}
	\maketitle 
	\textit{Introduction.---} Extending the concept of equilibrium phases of matter to non-equilibrium settings attracts perennial interest. Prominent examples involve many-body localization, where the spatial disorder enables eigenstate ordering even at high energy density, a phenomenon that is disallowed in thermal equilibrium~\cite{nandkishore2015many,abanin2019colloquium}. Periodically driven Floquet systems  enrich the zoo of non-equilibrium phases, where  exotic spatiotemporal behaviour can be realized, for instance, in discrete time crystals (DTCs)~\cite{khemani2016phase,else2016floquet,yao2017discrete} or Floquet topological phases~\cite{kitagawa2010topological,titum2016anomalous}.
	
	Here we ask: are there types of spatiotemporal ordering that are genuinely new to \textit{temporal} disorder, i.e.\ randomness in the drive, that lie beyond what can be achieved in conventional Floquet protocols? Closed time-dependent systems generally lack energy conservation~\cite{lazarides2014equilibrium}. Temporal disorder removes even the remaining quasi-conservation as in Floquet systems~\cite{abanin2015exponentially,kuwahara2016floquet}, opening up further deleterious energy absorption channels, generally believed to quickly heat up the system until all correlations become trivial and independent of the initial state. Therefore, unlike the spatial disorder which underpins eigenstate order~\cite{huse2013localization,pekker2014hilbert,chandran2014many,randall2021many},  temporal disorder generally diminishes interesting dynamical phenomenon~\cite{marino2012relaxation,rieder2018localization,long2021many,guarnieri2022time,zhao2022anomalous,mi2022time,bhattacharjee2022quasilocalization} and would, therefore, seem to hold little promise of exhibiting new spatiotemporal types of order. 
	
	Nonetheless, a transient but long-lived prethermal regime can exist if the heating rate is sufficiently controlled~\cite{else2017prethermal,weidinger2017floquet,weidinger2017floquet,dumitrescu2018logarithmically,else2020long,luitz2020prethermalization,mori2021heating,viebahn2021suppressing,fleckenstein2021thermalization,beatrez2021floquet,rubio2020floquet,collura2022discrete}. In the context of aperiodic driving, a low heating rate can specifically be realized in drive protocols such as random multipolar driving (RMD) or {hyperuniform driving with a suppressed low frequency driving spectrum}. Here, the correlated temporal disorder {may} lead to a tunably polynomially suppressed heating rate for fast drives~\cite{zhao2021random}.
	
	Prethermalization opens a long time window that can potentially hold a rich variety of non-equilibrium phases enabled by random drivings. Some cousins of Floquet phases, like DTCs and anomalous RMD insulators, have been recently discovered in randomly driven systems~\cite{zhao2021random,choudhury2021self,timms2021quantized,zhao2022anomalous,zheng2022anomalous}. However, none of them answer our question as the temporal disorder in these systems plays the role of an unwanted perturbation, which destabilizes their corresponding Floquet phases, albeit controllably gently. The challenge in moving beyond the Floquet paradigm is thus to employ temporal disorder sufficiently strong to qualitatively  modify the dynamical properties while sufficiently weak to  keep heating under control.
	
	In this work we provide an affirmative answer by presenting a prethermal phase characterized by two types of temporal correlations: conventional stroboscopic spatiotemporal DTC order coexists with temporally disordered micro-motion. This lies outside the established Floquet lore, where micro-motions are restricted to follow the stroboscopic time evolution via a deterministic gauge transformation~\cite{bukov2015universal}. We evade this constraint by randomly applying a spin-flip operation, leading to a nontrivial $\pi-$shifted Fourier spectrum of micro-motion that is distinct from that of the driving protocol. The stability of this spatiotemporal order can be analyzed via a Magnus expansion, and we derive a static effective Hamiltonian in the prethermal regime with a suppressed heating rate.
	
	In the following, we first introduce our temporally random driving protocol and elaborate on the prethermal time translation symmetry (TTS) breaking in a soluble instance. We then analytically investigate and numerically verify its stability away from solubility. Upon perturbing from solubility, the prethermal time scale grows algebraically with driving frequency. Remarkably, the scaling exponent strongly depends on the symmetry of the perturbations. Finally,  we show that the prethermal phase persists for generic initial states in a localized model.  {Beyond the example of RMD driving, we generalize our findings for random hyperuniform drivings}.
	
\begin{figure}
		\centering
		\includegraphics[width=\linewidth]{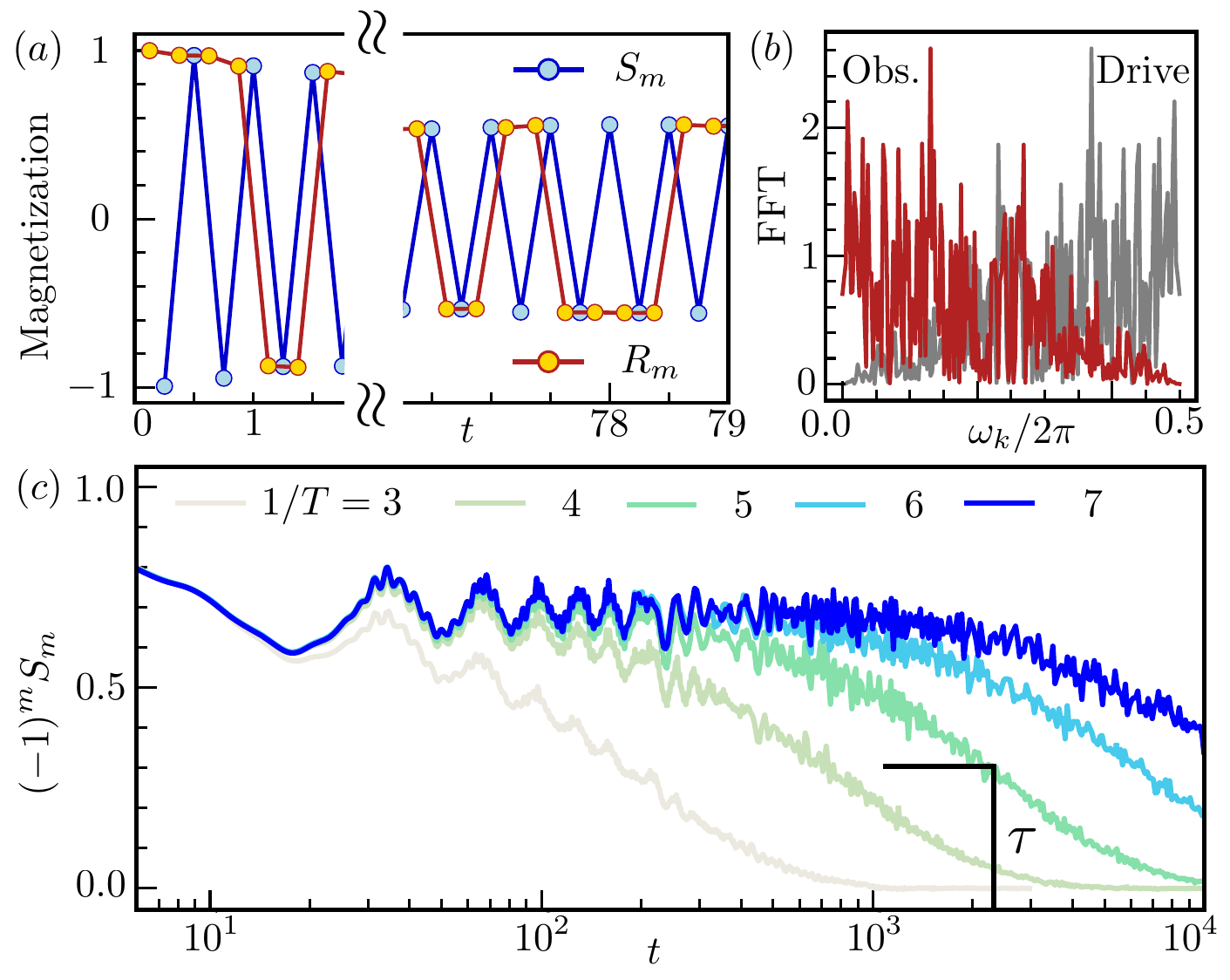}
		\caption{(a) Distinct temporal correlations emerge for micro-motions and stroboscopic times (for system size $L=26$). (b) Fourier modes of micro-motions exhibit a $\pi-$shifted spectrum different from the drive, defining a new type of TTS breaking. (c) This  persists for a long lifetime, which grows with driving frequency $1/T$. We use $J_z=1,J_x=0.1,J_y=0.2,B_z=0,\delta_r=0.15,L=18$.}
		\label{fig:dynamics}
	\end{figure}

	\textit{Driving protocol and soluble model.---}
	We consider a stepwise drive  with two  elementary time evolution operators for a spin$-1/2$ chain of length $L$,
	\begin{eqnarray}
		\begin{aligned}
			\label{eq:U0}
			&U_0^+ = U_zU_x, \ \ \ 
			U_0^- = U_xU_z,\\
			&U_z = \exp\left(-i{T}H_z/2\right),\ \ \  U_x = \exp\left(-i{T}H_x/2\right)
		\end{aligned}
	\end{eqnarray}
	where $H_z$  only involves the nearest-neighbor Ising interaction $H_z {=} \sum_jJ_z\sigma_{j}^z\sigma_{j+1}^z$ of strength $J_{z}$, and a field $H_x = B_x\sum_j\sigma_j^x$ of amplitude $B_x$. The soluble case is $B_x=\pi/T$, where the operator $U_x$ simplifies to the perfect global spin flip $X {=} \prod_j \sigma_j^x$ of the Ising $Z_2$ symmetry.
	As $H_z$ preserves this Ising symmetry, the product of two $U_0^{+}$ operators can be generated by the Hamiltonian $H_z$ as
	\begin{eqnarray}
		\label{eq:Z2symmetry}
		[U_0^{+}]^2 = \exp\big({-}iTH_z\big).
	\end{eqnarray}
	Consequently, for a Floquet drive generated by $U_0^+$ and for a $Z_2$ symmetry broken initial state, the local magnetization at stroboscopic times $mT$
	\begin{eqnarray}
		S_m = \sum_j \langle\sigma_j^z(0)\sigma_j^z(mT)\rangle/L,
	\end{eqnarray} exhibits period-doubling behavior
	with respect to the $T$ periodic spin flips, see the blue dots in Fig.~\ref{fig:dynamics}(a). Note, in this soluble case, the oscillation amplitude of $S_m$ never decays. Also, the micromotion is also strictly period doubled, with the magnetization, e.g., at half integer times  $(m+1/2)T$
	\begin{eqnarray}
		\label{eq.R_i}
		R_m = \sum_j\langle\sigma_j^z(0)\sigma_j^z(T/2+mT)\rangle/L,
	\end{eqnarray}
	behaving as that at integer times, as it is connected by the `gauge transformation' $U_x$.
	
	Now we remove the strict periodicity and randomly apply $U_0^{\pm}$ according to a random driving sequence $\{y_m\}$ where $y_m=\pm1$. Its discrete Fourier spectrum, $Y(\omega_k) =\sum_{m=0}^{M-1}y_m\exp\left(-i\omega_km\right)/\sqrt{M}$ where $M$ denotes the length of $\{y_m\}$,  exhibits a random distribution in frequency $\omega_k$ space with a flat envelope. A similar period doubling phenomenon still occurs at stroboscopic times as the relation in Eq.~\ref{eq:Z2symmetry} equally applies to any product of two operators  $U_0^{\pm}$. 
	
	However, the Floquet theorem does not apply and at half integer times, the magnetization is temporally disordered with the expression $R_m=(-1)^my_m$: its value depends on $y_m$ as $U_0^{\pm}$ determines whether spin-flip happens in the first or the second half of a period $T$; the phase $(-1)^m$ appears as the spin-flip changes the sign of magnetization and it disappears after even number of spin-flips. 
	The phase indeed implies a $\pi-$shifted spectrum $\widetilde{Y}(\omega_k) = {Y}(\omega_k+\pi)$, see proof in Supplementary Material (SM), but both of them follow the trivial and structureless frequency spectrum. 
	
	However, more interestingly, if ${Y}(\omega_k)$ is structured, this  $\pi-$shift generally leads to a different spectrum of local observables, hence generalizing the notion of prethermal TTS breaking to random driving protocols.
	We illustrate this idea by using the family of correlated random drives, the $n-$RMD protocol with a non-negative integer $n$ which quantifies the temporal correlation~\cite{zhao2021random}. $0-$RMD corresponds to the purely random case discussed above. Time evolution is generated by randomly applying one of the multipolar operators $U_n^{\pm}$, recursively defined as $U_n^{\pm}=U_{n-1}^{\mp}U_{n-1}^{\pm}$. For $n=1$, the envelop of the Fourier spectrum of the drive reads $Y_{1}(\omega_k) \sim \sqrt{1-\cos \omega_k}$ with a linear suppression at $\omega_k=0$, as shown by gray lines in Fig.~\ref{fig:dynamics} (b). The $\pi-$shift for the micro-motion persists and results in $\widetilde{Y}_{1}(\omega_k) \sim \sqrt{1+\cos \omega_k}$ where the suppression can be clearly observed at $\omega_k=\pi$ (red lines) in Fig.~\ref{fig:dynamics} (b).
	Indeed, this behavior exists for the full family of $n-$RMDs, where the envelop of the Fourier spectrum follows
	$
	Y_{n}(\omega_k) {\sim} \prod_{j=1}^{n}\sqrt{1-\cos \left(2^{j-1}\omega_k\right)},
	$ and a $\pi-$shift leads to 
	\begin{eqnarray}
		\begin{aligned}
			\widetilde{Y}_{n}(\omega_k)\sim\sqrt{1+\cos \omega_k}\prod_{j=2}^{n}\sqrt{1-\cos \left(2^{j-1}\omega_k\right)},
		\end{aligned}
	\end{eqnarray}
	for $n\geq2$. 
	Similar $\pi-$shifts should also occur for other random protocols as long as the driving sequence has a nontrivial frequency spectrum. 
	As a special feature of the n-RMD protocol, $\widetilde{Y}_{n}(\omega_k)$ is also a reflection of ${Y}_{n}(\omega_k)$, see Fig.~\ref{fig:dynamics}(b), which, however, is not a general property of TTS breaking in randomly driven systems.
	
	\textit{Stability.---} 
	A natural question is whether this TTS breaking persists away from the soluble case. The question of stability is also important for  experimental realizations. Naively, the answer would seem to be negative, simply because the continuous Fourier spectrum opens energy absorption channels to  destabilize the whole phenomenon. 
	
	However, we have previously shown that  generic many-body systems driven with an $n-$RMD protocol can exhibit algebraically long prethermal lifetimes in the high-frequency regime, \textit{i.e.}, the driving frequency is the dominant energy scale of the system~\cite{zhao2021random,mori2021rigorous}.
	By contrast, this is not the case here, as the spin-flip operator $U_x$ acts instantaneously regardless of driving frequency (or, alternatiely, requires the field strength to scale up with driving frequency as $B_x=\pi/T$), and the high-frequency condition is a priori {\it not} satisfied. Also, the fact the spin-flip operator occurs randomly in time prevents the existence of a deterministic rotating frame where the strong field always vanishes.
	
	We can nonetheless demonstrate that the period-doubling signal 
	persists for a finite but algebraically long lifetime even away from the fine-tuned point. We also find that the lifetime of the prethermal regime strongly depends on the symmetry of the perturbation. To do so, we consider two types of generic perturbations: (i) 
	the imperfection in spin flip operations with a field strength $B_x=\pi/T+\delta_r$ with $\delta_r$ always non-zero in the following discussion; and (ii) perturbations to the Hamiltonian $H_z$, which (a) either preserve the $Z_2$ symmetry, satisfying $X\Delta_{Z_2}X=\Delta_{Z_2}$ or (b) violate the symmetry $X\Delta X=-\Delta$. We can show that the concomitant prethermal lifetime scales as (a) $T^{-(2n+3)}$ and (b) $T^{-(2n-1)}$, respectively. 
	
	The multipolar operators $U_n^{\pm}$ being unitary, they can be written in the form $U_n^{\pm}=\exp[-i(2^{n-1}T)H^{\pm}_{n}]$ where $H^{\pm}_{n}$ is some static Hamiltonian operator. Such a Hamiltonian is generally nonlocal and hard to determine for many-body systems. However, by using a Magnus expansion,
	$H_n^{\pm}=\sum_{m=0}^{\infty}(2^{n-1}T)^m\Omega_{n,m}^{\pm}$, we can derive its high-frequency expansion that governs the prethermal dynamics~\cite{mori2021rigorous}. Subsequently we employ a Fermi's golden rule (FGR) argument to predict the scaling of the prethermal lifetime. 
	
	We start from the $Z_2$ preserving perturbations by changing $H_z$ in Eq.~\ref{eq:U0} to $H=H_z+\Delta_{Z_2}$. It leads to
	two dipolar operators 
	\begin{eqnarray}
		\begin{aligned}
			\label{eq:U1_Z2}
			U_1^- &= e^{-i{T}H/2}e^{-iT\delta_r \sum_i\sigma_i^x}e^{-i{T}H/2},\\
			U_1^+ &= e^{-i\frac{T}{2}\delta_r \sum_i\sigma_i^x}e^{-i{T}H} e^{-i\frac{T}{2}\delta_r \sum_i\sigma_i^x},
		\end{aligned}
	\end{eqnarray}
	where we use the $Z_2$ symmetry of the perturbed Hamiltonian $H$ and also the property $X^2=1$. Consequently, the strong field $B_x$ cancels out and the high-frequency regime is now well-defined if $1/T$ is much larger than any other local energy scale. 
	By treating $T$ as a small parameter, for $n=1$, the Magnus expansion leads to
	\begin{eqnarray}
		\label{eq.H_eff_1}
		H^{\pm}_{1} = H_z+\Delta_{Z_2}+\delta_r\sum_i\sigma_i^x+\mathcal{O}^{\pm}(T^2),
	\end{eqnarray}
	where the zeroth and the first order expression is the same for both $U_1^{\pm}$, with differing higher order terms  $\mathcal{O}^{\pm}(T^2)$ suppressed for high-frequency drives. Indeed,
	one can generalize this result (see SM) to larger $n$, and by induction show that the Hamiltonian, $H_{n}^{\mathrm{eff}}{\coloneqq}\sum_{m=0}^{n}(2^{n-1}T)^m\Omega_{n,m}^{\pm}$, truncated at $n-$th order is the same for $U_n^{\pm}$. 
	
	Therefore, the operator $H_{n}^{\mathrm{eff}}$ plays the role of a static effective Hamiltonian generated by an arbitrary sequence of $U_n^{\pm}$. For a generic non-integrable $H_{n}^{\mathrm{eff}}$, the system will first locally equilibrate to a prethermal state that can be locally captured by a Gibbs ensemble $\rho_{\mathrm{eff}}\sim \exp(-\beta_{\mathrm{eff}}H_{n}^{\mathrm{eff}})$. The effective inverse temperature $\beta_{\mathrm{eff}}$ can be determined by the initial expectation value of $H_{n}^{\mathrm{eff}}$. Additional higher order terms are generally different for $U_n^{\pm}$ and the most dominant ones' amplitude is $\mathcal{O}(T^{n+1})$. They appear randomly in time and induce a heating rate $\gamma\sim T\times (\mathcal{O}(T^{n+1}))^2\sim \mathcal{O}(T^{2n+3})$ according to FGR~\cite{mori2021rigorous}. Hence, the lifetime of the prethermal plateau should scale as $\tau\sim T^{-(2n+3)}$ which we numerically verify below. 
	
	Now we discuss the situation with $Z_2$ symmetry breaking perturbations with $H=H_z+\Delta$.
	For $n=1$, the same expression for $U_1^{-}$ is obtained as in Eq.~\ref{eq:U1_Z2}. However, $U_1^+$ is different
	\begin{eqnarray}
		U_1^+=  e^{-i\frac{T}{2}\delta_r \sum_i\sigma_i^x}e^{-i{T}(H_z-\Delta)} e^{-i\frac{T}{2}\delta_r \sum_i\sigma_i^x}.
	\end{eqnarray} A similar perturbation expansion leads to
	\begin{eqnarray}
		H^{\pm}_{1} = H_z\mp\Delta+\delta_r\sum_i\sigma_i^x+\mathcal{O}^{\pm}(T^2)\ .
	\end{eqnarray}
	Note the zeroth order terms differ by the symmetry breaking perturbation $\Delta$. Hence, for $n=1$ we do not have a well-defined time-independent Hamiltonian to approximate 1-RMD dynamics when the $Z_2$ symmetry of $H$ is explicitly broken. The existence of a prethermal regime is thus not guaranteed even for  large driving frequencies $1/T\to\infty$. 
	
	Prethermal behavior can, however, be obtained by increasing the multipolar order. For $n\geq2$, the perturbative expansion $H_{n}^{\mathrm{eff}}$ truncated at the $(n-2)$th order coincides for $U_n^{\pm}$. Consequently, it is the next, $(n-1)$st, order with amplitude $\mathcal{O}(T^{n-1})$ which destabilizes the system. The resulting prethermal lifetime scales as $\tau\sim T^{-(2n-1)}$. This scaling equally applies to $n=1$ if the symmetry breaking $\Delta$ is sufficiently weak, \textit{i.e.}, $\Delta$ does not strongly couple the initial low energy state and excited states of $H_z$. In the following, we  support our analysis with numerical simulation via exact diagonalization.
	
	\textit{Numerical simulation.---}
	We first consider an ordered initial state polarized in the positive $z-$direction, before extending the discussion to more general initial states. We consider a generic non-integrable Hamiltonian
	\begin{eqnarray}
		\label{eq:Hamiltonian_clean}
		H = H_z + \sum_j J_x\sigma_j^x\sigma_{j+1}^x+J_y\sigma_j^y\sigma_{j+1}^y+ B_z\sigma_j^z,
	\end{eqnarray}
	to generate $U_z$. Periodic boundary conditions are used
	such that the translation invariance permits us to simulate the
	dynamics for larger system sizes. The spin flip
	has the fixed rotation imperfection $\delta_r=0.15$. Non-zero $J_{x,y}$ introduces $Z_2$ preserving perturbations whereas $B_z$ violate the symmetry. 
	
	We first consider a single temporal disorder realization for 1-RMD with symmetry preserving perturbations ($B_z=0$). In Fig.~\ref{fig:dynamics}, we plot the magnetization $S_m$ at stroboscopic times as blue dots in panel (a).  Similarly to conventional discrete time crystals, $S_m$ oscillates with a period $2T$ with an amplitude decaying at short times but equilibrating at a non-zero value in the prethermal regime. The magnetization $R_m$ for the micro-motions (red) has approximately the same amplitude as $S_m$ but oscillates in a random fashion. The discrete Fourier spectrum of  $R_m/|R_m|$ up to the time $t=100$ is plotted in Fig.~\ref{fig:dynamics}(b). A clear suppression occurs at $\omega_k=\pi$ as predicted whereas the drive (gray) has a suppression at $\omega_k=0$, suggesting that the new type of TTS breaking survives perturbations in the prethermal regime. 
	
	\begin{figure}
		\centering
		\includegraphics[width=\linewidth]{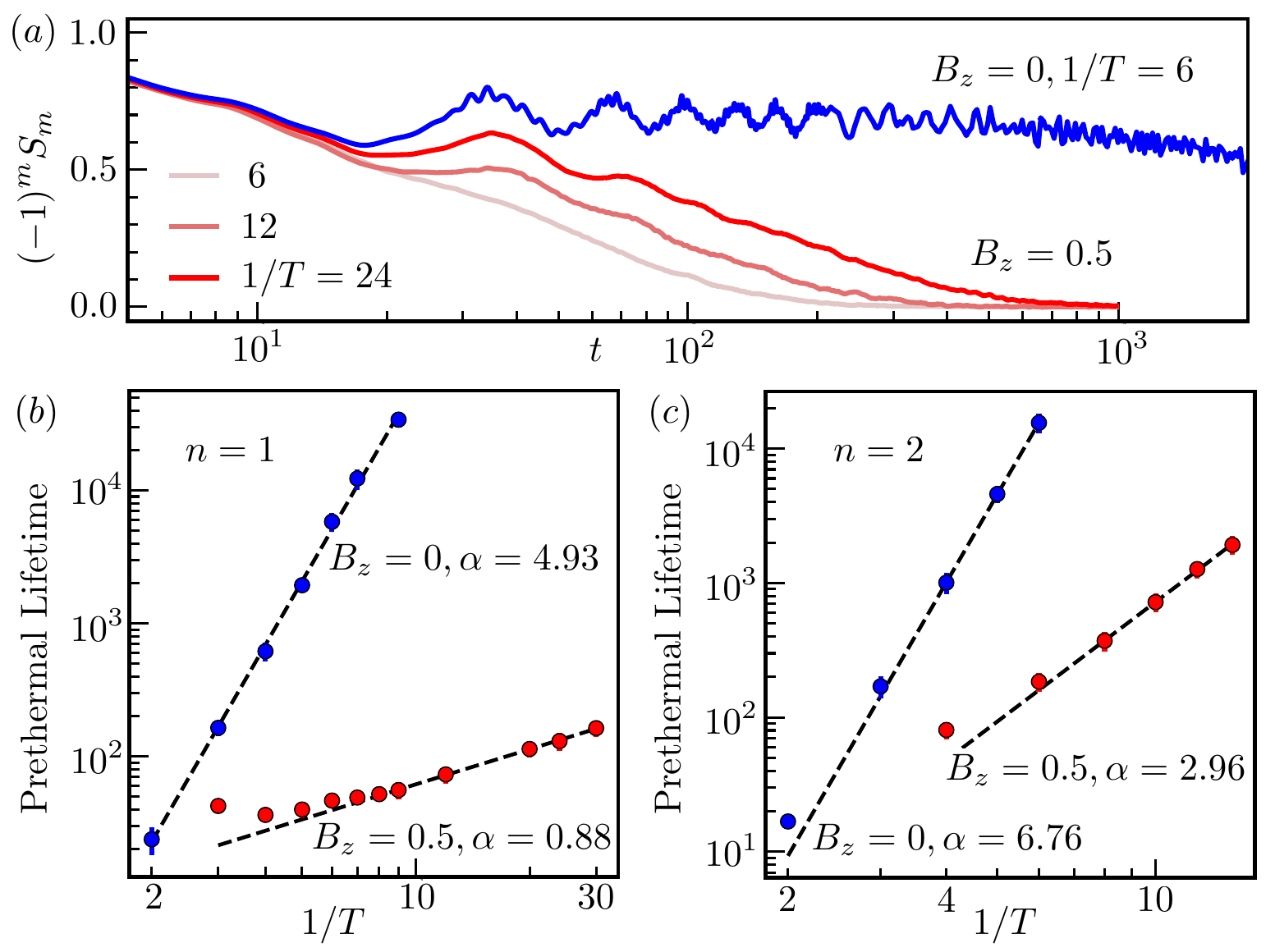}
		\caption{Prethermal lifetime $\tau$ strongly depends on the presence/absence (blue/red lines) of $Z_2$ symmetry in the perturbation. (a) Dynamics of the stroboscopic magnetization. (b) Algebraic lifetime scaling versus driving frequency $1/T$ for multipolar order $n=1$. Scaling exponent is approximately $2n+3$ for $Z_2$ preserving perturbation with $B_z=0$, and $2n-1$ for $B_z=0.5$ for $n=1$. (c) Scaling results for $n=2$.  We use $J_z=1,J_x=0.1,J_y=0.2,B_z=0,\delta_r=0.15,L=18$.}
		\label{fig:scaling}
	\end{figure}
	In Fig.~\ref{fig:dynamics}(c), after averaging over 200 temporally disordered realizations, $(-1)^mS_m$ is plotted and different colors correspond to different driving frequencies. After a short transient period, it relaxes to a non-zero value. The system heats up to infinite temperature after a long time scale, which increases for larger driving frequencies.
	A similar phenomenon also occurs when $Z_2$ symmetry is broken ($B_z\neq 0$) as shown in Fig.~\ref{fig:scaling} (a). But for a fixed frequency, \textit{e.g.}, $1/T=6$ (light red), $(-1)^mS_m$ decays much faster than the $Z_2$ preserving case (blue), highlighting the importance of symmetry in stabilizing the non-equilibrium phases with random drives.
	
	The dependence of the prethermal lifetime can be further quantified by setting a threshold value $s_0$ for the magnetization. We extract the time $t_{s_0}$ where $(-1)^mS_m$ first drops below $s_0$, and the prethermal lifetime $\tau$ is determined as the average $\langle t_{s_0}\rangle_{s_0}$ for five different threshold values $0.32,0.32\pm0.05,0.32\pm0.025$. The average is performed to reduce numerical noise and the following results do not rely on specific threshold values. In Fig.~\ref{fig:scaling} (b), $\tau$ is plotted with the error bar denoting the standard deviation and both axes use a log scale. A linear dependence is observed for both types of perturbations, suggesting an algebraic scaling $\tau\sim (1/T)^{\alpha}$. The exponent $\alpha$ is obtained by a linear fit in the high frequency regime, and we obtain the scaling exponent $\alpha\approx2n+3$ for $B_z=0$ and $\alpha\approx2n-1$ for $B_z\neq 0$. This scaling exponent is tunable by increasing the multipolar order as we verify for $n=2$ in Fig.~\ref{fig:scaling} (c).
	
	\begin{figure}
		\centering
		\includegraphics[width=\linewidth]{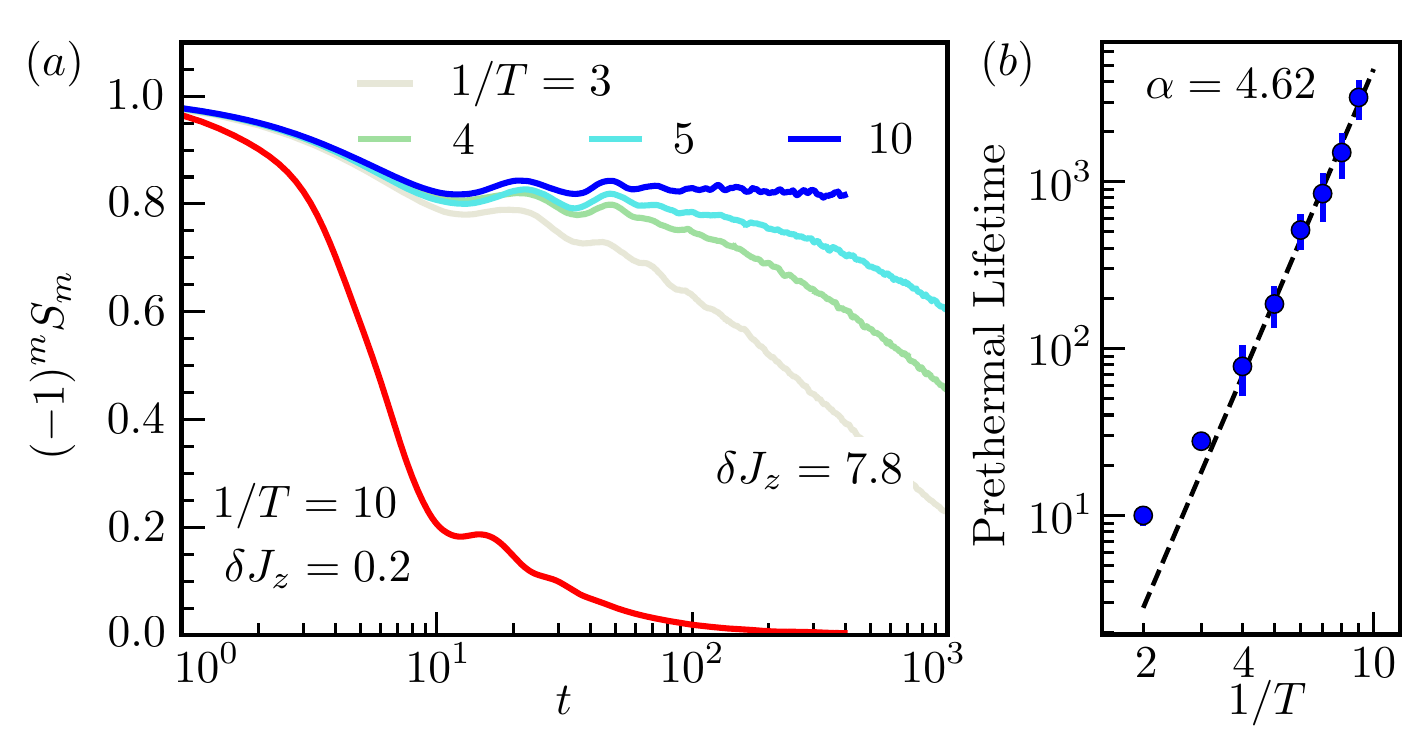}
		\caption{(a) Dynamics for random product state in $z-$direction. Prethermal phases persist for sufficiently strong disorder with a lifetime increasing for larger driving frequencies. For weak disorder (red), the spatiotemporal order quickly vanishes. (b) Algebraic lifetime scaling versus frequency with exponent close to $2n+3$. We use $J_z=1,J_x=0.1,B_z=0,\delta_r=0.08,n=1,L=16$.}
		\label{fig:MBL}
	\end{figure}
	
	The persistence of the spatiotemporal order relies on the fact that the polarized initial state corresponds to a sufficiently low temperature of the effective Hamiltonian $H_{n}^{\mathrm{eff}}$ in the prethermal regime~\cite{pizzi2021classical}. For more generic initial states at a finite temperature, $(-1)^mS_m$ quickly drops to zero and this prethermal phase will not exist, in accordance with the absence of long range order at finite temperature in one-dimension. This issue can be resolved by introducing, for instance, sufficiently strong spatial disorder to realize the eigenstate order even at high temperatures. 
	
	Let us thus consider the Hamiltonian $
	H = \sum_j(J_z+J_j)\sigma_j^z\sigma_{j+1}^z + J_x\sigma_j^x\sigma_{j+1}^x,
	$ with the spatially disordered couplings $J_j$ randomly chosen from $[-\delta J_z/2,\delta J_z/2]$. The system starts from a product state containing spins polarized randomly in the $\pm z$ direction with total magnetization zero. We perform 400 simulations to average over different initial states, spatial and temporal disorder realizations with the multipolar order $n=1$. The magnetization $(-1)^mS_m$ is plotted in Fig.~\ref{fig:MBL} (a). For sufficiently strong disorder $\delta J_z=7.8$ (blue), with localization established, the system maintains strong memory of the initial state, $(-1)^mS_m\approx 0.8$, in the prethermal regime. We extract the prethermal lifetime when the magnetization starts deviating from the prethermal plateau by using five different threshold values $0.76, 0.76\pm 0.02,0.76\pm 0.01$, and plot it in Fig.~\ref{fig:MBL} (b) where an algebraic scaling is observed. We do not observe notable finite size effects in our simulation, see details in SM. The fitted scaling exponent is close to $2n+3$ as predicted. In contrast, for a weak disorder $\delta J_z=0.2$ (red), eigenstates of the truncated effective Hamiltonian are not ordered at a high energy density, hence, the
	system quickly heats up  with a vanishing magnetization.
	
	\textit{Discussion.---}
	\begin{figure}
		\centering
		\includegraphics[width=\linewidth]{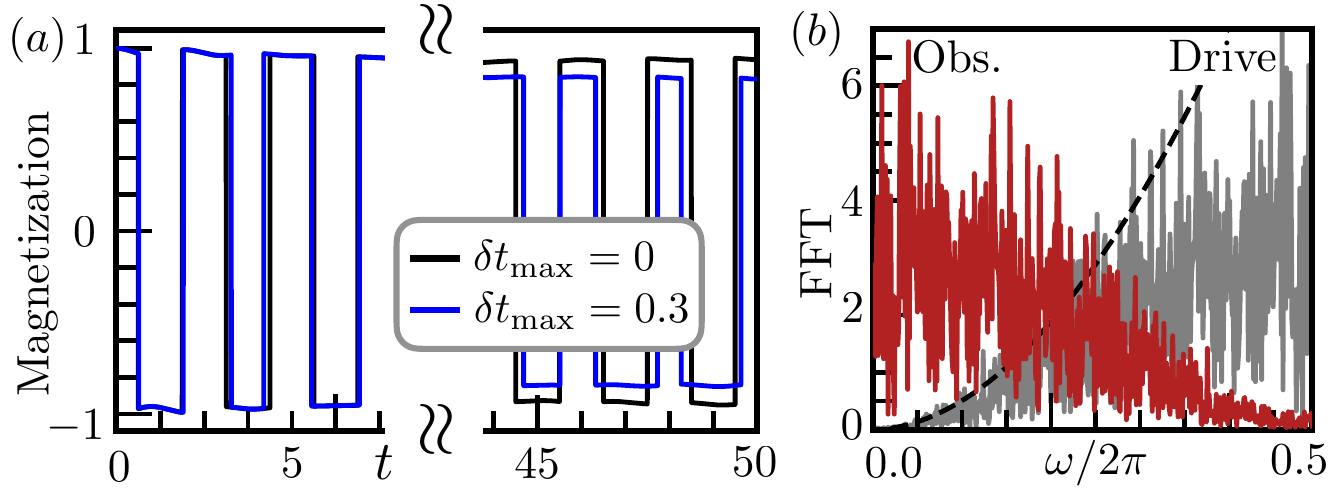}
		\caption{(a) Dynamics of the magnetization. For blue line, spins flip randomly according to a hyperuniform sequence, while for black line spins flip deterministically. System size $L=20$ and rotation imperfection $\delta'_r=0.01.$  (b) Fourier modes of magnetization exhibit a
			$\pi-$shifted spectrum different from the drive. The algebraic suppression (dashed black) has an exponent $\alpha/2$. We use $J_z=1,J_x=0.2,\alpha=3.8,L=16,\delta'_r=0,\delta t_{\mathrm{max}}=0.1, T=1$ for the numerical simulation.}
		\label{fig:hyper}
	\end{figure}
	We show that a new type of spatiotemporal order absent in Floquet systems can be achieved by structured random protocols, which is manifest through a $\pi-$shifted spectrum of the dynamics of local observables compared to the spectrum of the drive. Many-body localization is employed here to stabilize the $Z_2$ ordering in the prethermal regime. It can be alternatively achieved by using higher dimensional spin models or long-range interactions in 1D~\cite{machado2020long,kyprianidis2021observation,pizzi2021classical,ye2021floquet}, and classical spin models will also be suitable for large scale numerical simulations~\cite{howell2019asymptotic,yue2022space}.
	
	{We emphasize that, although our results build on the concrete $n-$RMD protocol, this novel type of TTS breaking is quite generic. To provide another concrete example, consider a driving protocol with the Hamiltonian $H=H_z + \sum_j J_x\sigma_j^x\sigma_{j+1}^x$ to generate the time evolution and apply $U_x=\exp(-i(1+\delta'_r)\pi/2)$ with imperfection $\delta'_r$ to flip the spins instantaneously at time $t=mT+T/2+\delta t_m$ for integer $m$ and random $\delta t_m\in [\delta t_{\mathrm{max}},\delta t_{\mathrm{max}}]$. This randomness can be chosen to be ``hyperuniform", i.e.\ with suppressed large-scale fluctuations~\cite{torquato2018hyperuniform,crowley2019quantum}, leading to an algebraic suppression of low frequencies similar to the RMD sequence but with a continuously tunable scaling exponent $\alpha/2$ , see details in the SM. In Fig.~\ref{fig:hyper} (a), starting from the initial state $\prod_i\ket{\uparrow}$, we plot the time evolution of the magnetization $\sum_j\langle \sigma_j^z\rangle/L $ where the regular period-doubling behavior occurs stroboscopically. In contrast, the random micromotions exhibit the novel $\pi-$shifted spectrum as shown in Fig.~\ref{fig:hyper} (b). Such a pattern can also be long-lived by choosing a small $\delta t_{\mathrm{max}}$, see SM. A systematic investigation of the prethermal lifetime and its relation to hyperuniformity is an intriguing subject for future study.}
	
	Another important feature is the symmetry dependence of the prethermal lifetime scaling. An intriguing direction for the future is then a systematic symmetry classification of heating dynamics. Similar questions are also worth studying in Floquet systems and quasi-periodically driven systems~\cite{verdeny2016quasi,nathan2020quasiperiodic,friedman2020topological,long2021many,zheng2022anomalous}. 
	
	Going forward, we anticipate the extension of our protocol to Floquet topological phases and their generalizations~\cite{wampler2022stirring,dumitrescu2022dynamical}. In Ref.~\cite{titum2016anomalous}, the soluble limit involves four-step hopping processes and a fifth-step random disorder potential. One can reshuffle the fifth step, such that particle hopping is temporally disordered while stroboscopic dynamics remain deterministic. Investigating the prethermal phenomenon away from this soluble limit will be worth pursuing in the future.
	
	\textit{Acknowledgements.---} This work is in part supported by the Deutsche Forschungsgemeinschaft  under  cluster of excellence ct.qmat (EXC 2147, project-id 390858490). The research is
	part of the Munich Quantum Valley, which is supported by
	the Bavarian state government with funds from the
	Hightech Agenda Bayern Plus.
	
	\bibliography{Reference}
	\appendix
	\newpage~
	\section{$\pi-$shifted Fourier spectrum for RMD}
	Suppose all spins are flipped either in the first ($U_0^-$) or the second half ($U_0^+$) of a single period $T$ according to a random string, e.g., $\{y_m\}=\{+1,-1,-1,-1,+1,\dots\}$ which contains $M$ number of $\pm1$. Its discrete Fourier decomposition, defined as
	\begin{eqnarray}
		Y(x_k) = \frac{1}{\sqrt{M}}\sum_{m=0}^{M-1}y_m\exp\left(-i2\pi x_km\right),
	\end{eqnarray}
	where $x_k=k/N\in[0,1)$ for $k\in[0,N-1)$, exhibits certain nontrivial distributions, for instance, a dipolar structure as shown in Fig.~\ref{fig:dynamics}(b). $Y(x_k)$ is obtained by fast Fourier transformation and plotted versus $x_k$. Only half of the $x_k$ is plotted, and the other half is symmetric with respect to $x_k=1/2$ as the string elements are all real.
	Consider a single spin $\ket{\uparrow}$ as the initial state, the magnetization in z-direction at half integer times reads $\{\widetilde{y}_m\}=\{+1,+1,-1,+1,+1,\dots\}$, which differ from the driving $\{{y}_m\}$ by a phase $(-1)^m$ as the spin comes back to its initial polarization after two spin flips. It leads to a new Fourier spectrum
	\begin{eqnarray}
		\begin{aligned}
			\widetilde{Y}(x_k) &= \frac{1}{\sqrt{M}}\sum_{m=0}^{M-1}\widetilde{y}_m\exp\left(-i2\pi x_km\right)\\
			&= \frac{1}{\sqrt{M}}\sum_{m=0}^{M-1}(-1)^m{y}_m\exp\left(-i2\pi x_km\right)\\
			&= \frac{1}{\sqrt{M}}\sum_{m=0}^{M-1}{y}_m\exp\left[-i2\pi m(x_k+1/2)\right],
		\end{aligned}
	\end{eqnarray}
	resulting in the relation $\widetilde{Y}(x_k) = {Y}(x_k+1/2)$. Alternatively, in term of frequencies $\omega_k=2\pi x_k$, it implies a $\pi-$shifted spectrum $\widetilde{Y}(\omega_k) = {Y}(\omega_k+\pi)$ with $\omega_k\in[0,2\pi)$.
	Specifically, for $n-$RMD drive, the envelop of the Fourier spectrum of the drive has been shown to be
	\begin{equation}
		\label{eq.RMDspectrum}
		Y_{n}(\omega_k) \sim \prod_{j=1}^{n}\sqrt{1-\cos \left(2^{j-1}\omega_k\right)},
	\end{equation} and a $\pi-$shift leads to 
	\begin{eqnarray}
		\begin{aligned}
			\label{eq:pi-shift-RMD}
			&\widetilde{Y}_{n}(\omega_k) \sim \prod_{j=1}^{n}\sqrt{1-\cos \left(2^{j-1}\omega_k+2^{j-1}\pi\right)}\\
			&= \sqrt{1+\cos \omega_k}\prod_{j=1}^{n-1}\sqrt{1-\cos \left(2^{j}\omega_k\right)},
		\end{aligned}
	\end{eqnarray}
	where a frequency suppression occurs at $\omega_k=\pi$, as shown in Fig.~\ref{fig:dynamics}(b). Another interesting property in Fig.~\ref{fig:dynamics}(b) is observable spectrum is a reflection of the driving spectrum with respect to $\omega_k=\pi/2$. This generally happens for $n-$RMD protocols as 
	\begin{eqnarray}
		\begin{aligned}
			\label{eq.reflection}
			\widetilde{Y}_n(\pi/2+x) &= {Y}_n(-\pi+\pi/2+x) \\
			&={Y}_n(-\pi/2+x) = {Y}_n(\pi/2-x),
		\end{aligned}
	\end{eqnarray}
	where the last equal sign can be verified from Eq.~\ref{eq.RMDspectrum}. However, such reflection is not a general property for other types of spectrum.
	
	\section{Derivation of the effective Hamiltonian}
	For $Z_2$ symmetry preserving perturbation, we  show that
	the effective Hamiltonian $H_{n}^{\mathrm{eff}}=\sum_{m=0}^{n}(2^{n-1}T)^m\Omega_{n,m}^{\pm}$ truncated up to the $n-$th order are same for $U_{n}^{\pm}$. In Eq.~\ref{eq.H_eff_1} we have already shown that it works for $n=1$.
	Let us first follow Ref.~\cite{mori2021rigorous} and prove for each $n$, we have
	\begin{equation}
		\Omega_{n,m}^{+}=\Omega_{n,m}^{-}=2^{-m(n-m)}\frac{\Omega_{m,m}^{+}+\Omega_{m,m}^{-}}{2}.
		\label{eq:Magnus_formula}
	\end{equation}
	for all $m\leq n$ by induction. We first assume Eq.~\ref{eq:Magnus_formula} is true, then $U_{n+1}^{\pm}=U_{n}^{\mp}U_{n}^{\pm}$ reduces to
	\begin{align}
		&U_{n+1}^{\pm}\nonumber\\&{=}\exp\left[{-i}\left(H^\mathrm{eff}_{n}(2^{n-1}T){+}\sum_{m=n+1}^\infty\Omega_{n,m}^{\mp}(2^{n-1}T)^{m+1}\right)\right]
		\nonumber \\
		&\times\exp\left[-i\left(H^\mathrm{eff}_{n}(2^{n-1}T)+\sum_{m=n+1}^\infty\Omega_{n,m}^{\pm}(2^{n-1}T)^{m+1}\right)\right].
	\end{align}
	This can be rewritten as a single exponential by using the Baker-Campbell-Hausdorff formula. 
	\begin{eqnarray}
		\begin{aligned}
			U_{n+1}^{\pm}&=\exp\Big[-i\big(H^\mathrm{eff}_{n}2^{n}T
			+\big(\Omega_{n,n+1}^{+}+\Omega_{n,n+1}^{-}\big)\\
			&\times(2^{n-1}T)^{n+2}
			+\mathcal{O}\left(T^{n+3}\right)\big)\Big].
		\end{aligned}
	\end{eqnarray}
	By comparing it with the direct Magnus expansion of
	$U_{n+1}^{\pm}=\exp\left[-i\sum_{m=0}^\infty(2^{n}T)^{m+1}\Omega_{n+1,m}^{\pm}\right]$, we obtain
	\begin{eqnarray}
		\begin{aligned}
			&\Omega_{n+1,m}^{+}=\Omega_{n+1,m}^{-}=2^{-m}\frac{\Omega_{n,m}^{+}+\Omega_{n,m}^{-}}{2}\\&=2^{-m(n+1-m)}\frac{\Omega_{m,m}^{+}+\Omega_{m,m}^{-}}{2}
			\quad \text{for all }m\leq n+1,
			\label{eq:formula}
		\end{aligned}
	\end{eqnarray}
	which shows that Eq.~(\ref{eq:Magnus_formula}) also holds for $n+1$. Therefore, by induction we conclude that Eq.~(\ref{eq:Magnus_formula}) holds for every $n$. Note, this scaling is different from our previous result in Ref.~\cite{zhao2021random} where Eq.~\ref{eq:Magnus_formula} works for $m\leq n-1$. This happens because $U_0^{\pm}$ employed in this work already has a dipolar structure according to the definition in Ref.~\cite{zhao2021random}.
	
	Similar argument can be carried out for symmetry breaking perturbation. However, Eq.~\ref{eq:Magnus_formula} only holds true for all $m\leq n-2$. One can easily verify this for $n=2$ where the $\pm\Delta$ term cancels out in the lowest order effective Hamiltonian, leading to  $H_z+\delta_r\sum_i\sigma_i^x$.
	
	\section{Simulation for different system sizes}
	
	\begin{figure}
		\centering
		\includegraphics[width=\linewidth]{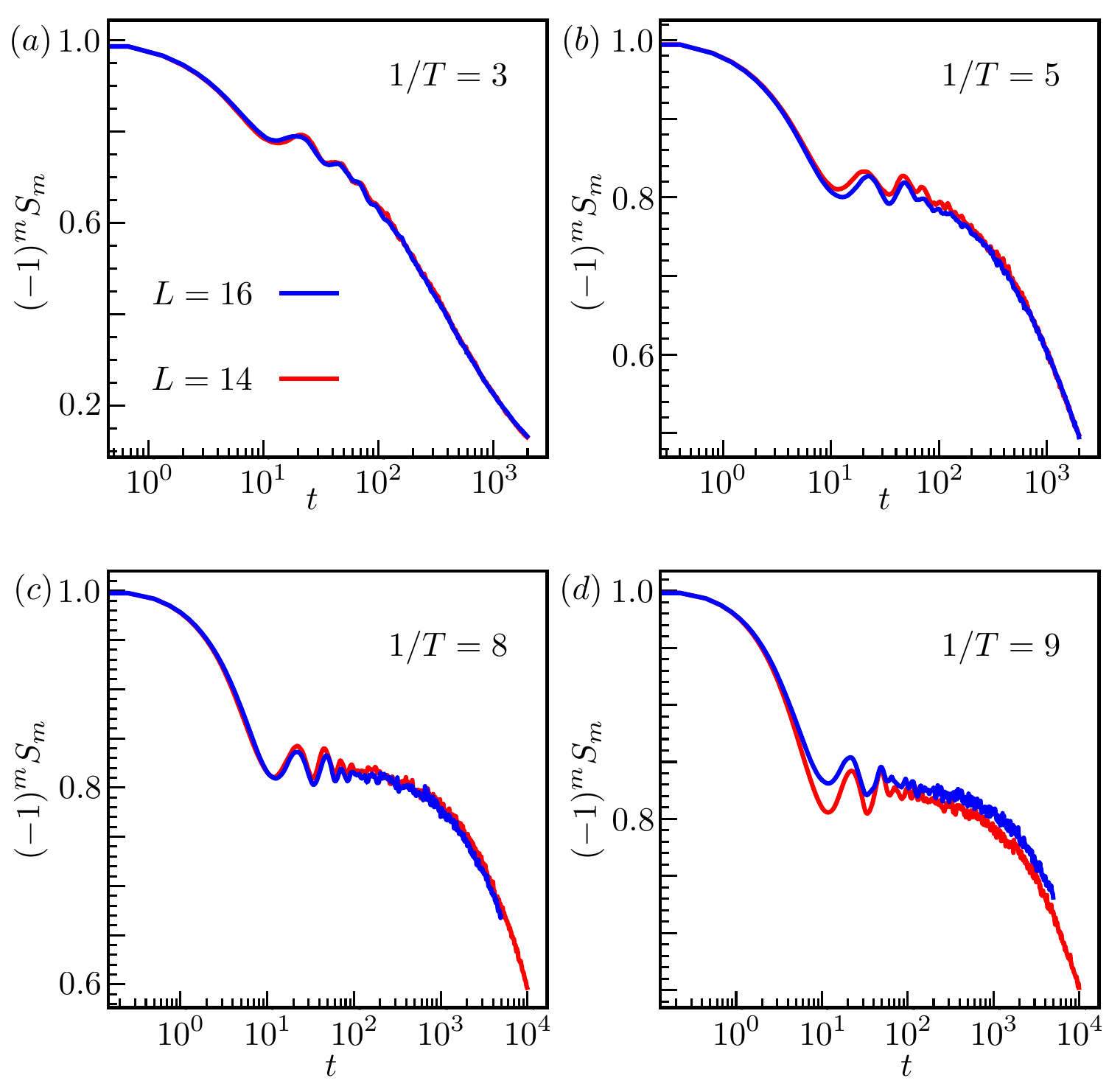}
		\caption{Dynamics of magnetization for different driving frequencies and different system sizes. We use $J_z=1,J_x=0.1,B_z=0,\delta_r=0.08,n=1,L=16$ for numerical simulation.}
		\label{fig:MBL_dynamics_size}
	\end{figure}
	We plot the dynamics of the localized model for different system sizes and different driving frequencies in Fig.~\ref{fig:MBL_dynamics_size}. Initial states are chosen as product states with randomly polarized spins in $z$ direction, in the sector with zero total magnetization. Results have been averaged over 400 simulations with different initial states, spatial and temporal disorder. We notice that for small driving frequencies where the system quickly heats up, for instance, $1/T=3$, results converge rapidly with a small number of simulations. While for larger driving frequencies, more simulations are needed to average out the fluctuation between different disorder realizations. 
	
	\begin{figure}
		\centering
		\includegraphics[width=0.55\linewidth]{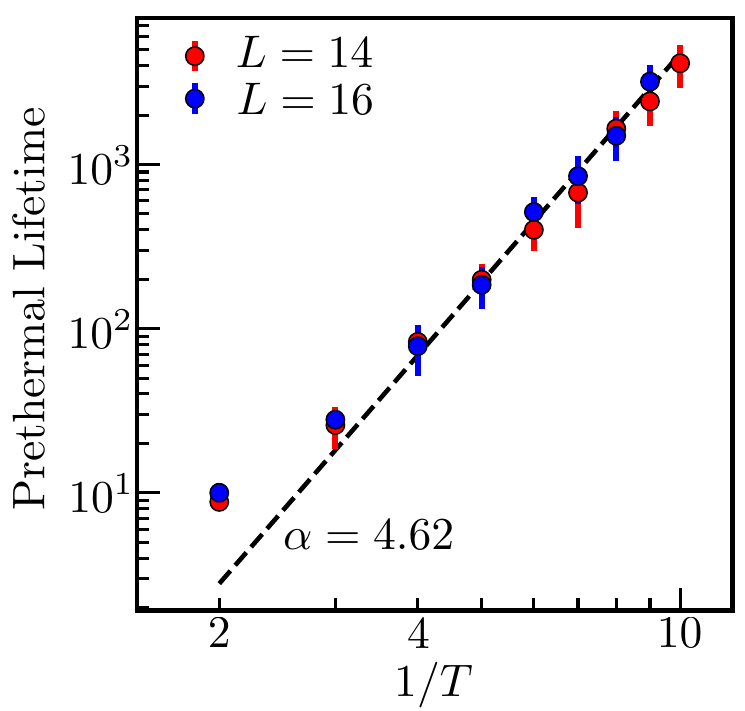}
		\caption{Scaling for different system sizes}
		\label{fig:MBL_scaling_size}
	\end{figure}
	
	The averaged results are very similar for $L=14$ and $L=16$ without notable finite size effects in the simulated parameter regimes and time scales. A similar dependence of the prethermal lifetime on the driving frequencies is observed, as shown in Fig.~\ref{fig:MBL_scaling_size}. Note, finite size effect may still occur for very high frequency driving and at late times, for instance, $t \gtrsim10^4$, which goes beyond what we can simulate for random multipolar drivings.

	\subsection{Hyperuniform protocol and its $\pi-$shifted Fourier spectrum}
	{Following Ref.~\cite{crowley2019quantum}, 
		a hyperuniform driving has suppressed low frequency components in its Fourier spectrum. Concretely, one can consider 
		\begin{eqnarray}
			q_n=\frac{1}{\sqrt{L}} \sum_{k} q_{k} e^{-i k n},
		\end{eqnarray}
		with $k=2 \pi m / L$, $m=1 \ldots L$
		and $q_k$ are random numbers with correlation 
		\begin{equation}
			\label{eq.correlation}
			S_{\alpha}\left(k, k^{\prime}\right) \equiv\left[q_{k} q_{-k^{\prime}}\right] \sim|k|^{\alpha} \delta_{k k^{\prime}},
		\end{equation}
		where $\alpha$ is the hyperuniformity parameter to quantify the low frequency suppression
		$[\cdot]$ denotes random average. For our discrete RMD protocol $\alpha=2n$, but here in general $\alpha$ can be tuned continuously. For instance, we can use
		\begin{equation}
			\label{eq.core}
			q_{k}=|\sin (k / 2)|^{\alpha / 2} \frac{1}{\sqrt{L}} \sum_{j} \xi_{j} e^{i k j}, \ \text{for}\ j=1 \ldots L, 
		\end{equation}
		where $\xi_{j}$ are randomly drawn from $\pm1$, and one can perform inverse Fourier transformation to obtain $q_j$.
		The correlation Eq.~\ref{eq.correlation} results in an algebraic suppression at small frequency of the Fourier spectrum of $q_j$ with exponent $\alpha/2$. }
	
	{Now we introduce the following driving protocol exhibiting hyperuniformity. We first consider the piece-wise constant function that is non-vanishing only in the time interval $[nT,(n+1)T]$:
		\begin{eqnarray}
			h_n(t):= \begin{cases}1, & nT\le t<nT+T/2+\delta t_n \\ -1, & nT+T/2+\delta t_n\le t<(n+1)T\\
				0,  &\text{else}\end{cases}
		\end{eqnarray} 
		where the temporal fluctuation  $\delta t_n\in [-\delta t_{\mathrm{max}},\delta t_{\mathrm{max}}]$, sampled from a hyperuniform sequence, namely $t_n=\delta_{\mathrm{max}}q_n$.
		The overall driving function is 
		\begin{eqnarray}
			{h}(t)=\sum_nh_n(t),
		\end{eqnarray}
		and if $h(t)=1$, the Hamiltonian $H=\sum_j  J_z\sigma_j^z\sigma_{j+1}^z+J_x\sigma_j^x\sigma_{j+1}^x$, is applied on the many-body spin chain; if $h(t)=-1$, we first apply $U_x=\exp(-i(1+\delta'_r)\pi/2)$ instantaneously then again use $H$ to continue the time evolution. }
	
	\begin{figure}[h]
		\centering
		\includegraphics[width=0.6\linewidth]{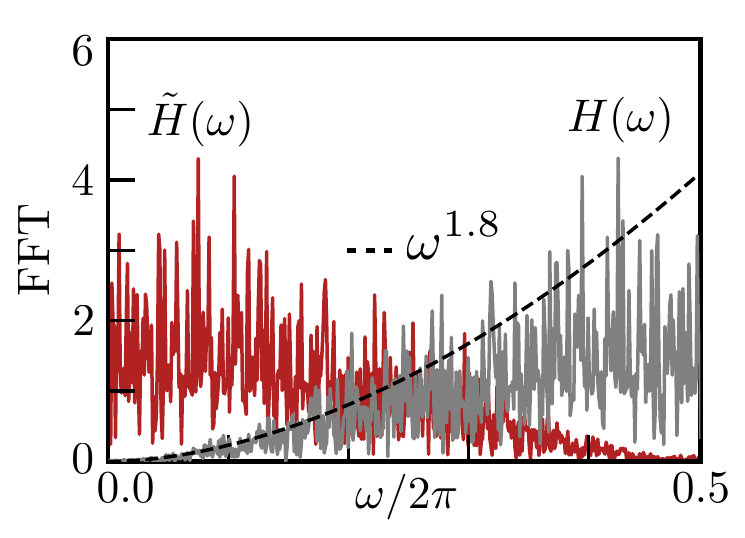}
		\caption{Comparison between the driving protocol and the magnetization at the soluble limit. We use $\alpha=3.6$ and $T=1$ to generate the hyperuniform driving function so the frequency suppression has the scaling exponent $\alpha/2=1.8$. The black dashed line represents a function with scaling $\omega^{\alpha/2}$ which correctly capture the envelop of the driving spectrum especially for $\omega\to 0.$ $\pi-$shift happens to the magnetization spectrum so the suppression happens for $\omega/2\pi\to0.5.$ }
		\label{fig:fft_hyper}
	\end{figure}
	
	{The hyperuniformity naturally arises in the Fourier spectrum of $h(t)-h'(t)$ where $h'(t)$ is the periodic counterpart of $h(t)$, i.e., $\delta t_{\mathrm{max}}=0.$ Explicitly,
		\begin{eqnarray}
			\begin{aligned}
				{H}(\omega)&=\int_{-\infty}^{\infty}e^{-i\omega t}(h(t)-h'(t))dt\\
				&=e^{-\omega T/2}\sum_n  e^{-i\omega n T}(e^{-i\omega \delta t_n}-1)/(-i\omega),
			\end{aligned}
		\end{eqnarray}
		and for $\delta t_{\mathrm{max}}\ll 1$ the equation above reduces to  
		\begin{eqnarray}
			\begin{aligned}
				{H}(\omega)=e^{-\omega T/2}\sum_n e^{-in\omega T} \delta t_n,
			\end{aligned}
		\end{eqnarray}
		exhibiting frequency suppression for $\omega\to 0$ with the scaling exponent $\alpha/2$, the same as $q_k$.}
	
	\begin{figure}[h]
		\centering
		\includegraphics[width=\linewidth]{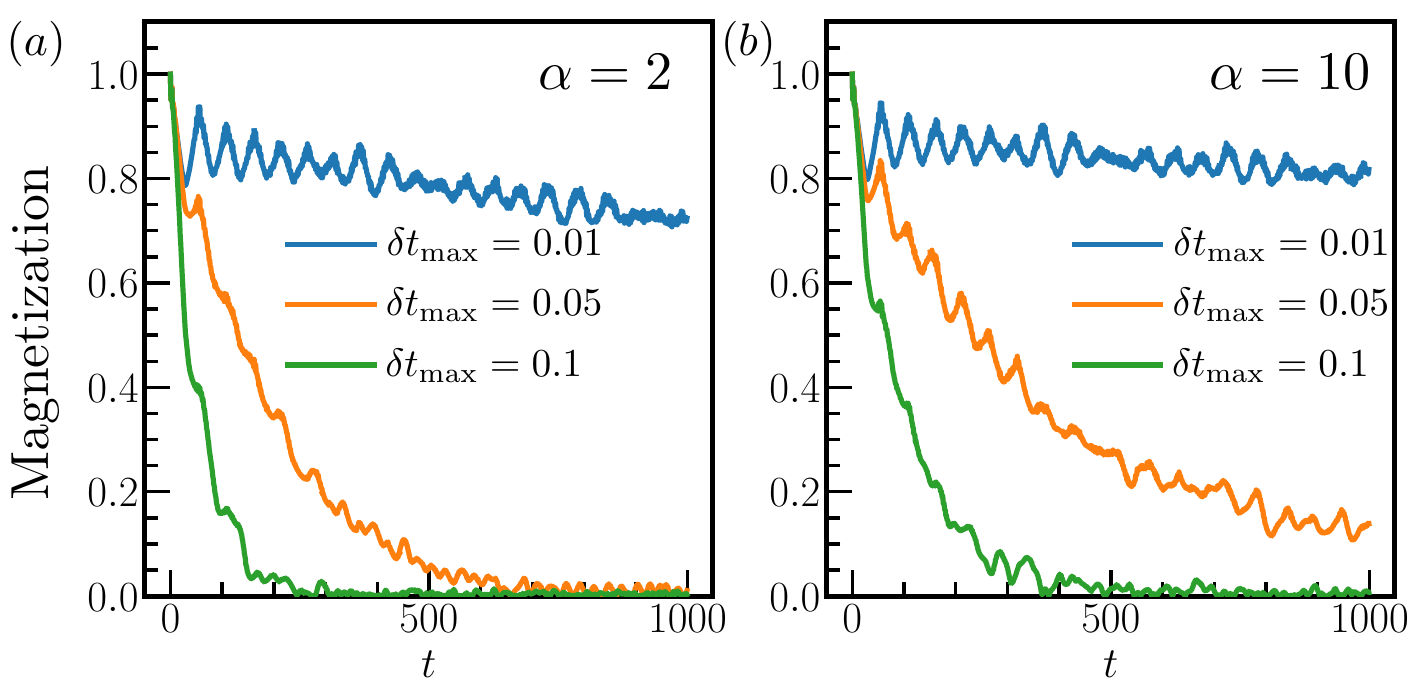}
		\caption{Dynamics of magnetization for different $\delta t_{\mathrm{max}}$. Heating rates can be suppressed via a smaller $\delta t_{\mathrm{max}}$ and larger hyperuniform parameter $\alpha$.   We use $J_z=1,J_x=0,L=16,\delta_r=0.1,T=1$ for the numerical simulation.}
		\label{fig:hyper-SM}
	\end{figure}
	{At the soluble limit, namely the spin flip $U_x$ is perfect ($\delta'_r=0$), $J_x=0$, and the initial state is $\ket{\uparrow}$  the z-magnetization evolves in time as
		\begin{eqnarray}
			\tilde{h}(t)=\sum_nh_n(t)(-1)^n,
		\end{eqnarray}
		where $(-1)^n$ appears due to the spin flip operation.
		The Fourier spectrum reads
		\begin{eqnarray}
			\begin{aligned}
				\Tilde{H}(\omega)&=\int_{-\infty}^{\infty}e^{-i\omega t}(\tilde{h}(t)-\tilde{h}'(t))dt\\
				&=e^{-\omega T/2}\sum_n e^{-i\pi n} e^{-i\omega n T}(e^{-i\omega \delta t_n}-1)/(-i\omega),
			\end{aligned}
		\end{eqnarray}
		where we use $(-1)^n=e^{-i\pi n}$. Suppose $\delta t_{\mathrm{max}}\ll 1$, the equation above reduces to  
		\begin{eqnarray}
			\begin{aligned}
				\Tilde{H}(\omega)=e^{-\omega T/2}\sum_n e^{-in(\omega T+\pi)} \delta t_n.
			\end{aligned}
		\end{eqnarray}
		Remarkably, it suggests a $\pi-$shift is happening in the spectrum, i.e., $|\tilde{H}(\omega)|= |H(\omega+\pi/T)|$, which is confirmed  for the soluble case in Fig.~\ref{fig:fft_hyper}.}

	{Now we simulate the time evolution numerically for a spin chain of length $L=16$ with different hyperuniform parameter, $\alpha=2$ and 10 in Fig.~\ref{fig:hyper-SM} (a) and (b), respectively. The absolute value of the z-magnetization at stroboscopic times $t=mT$ is plotted. In both cases, a smaller maximum temporal fluctuation $\delta t_{\mathrm{max}}$ generally enlarges the lifetime of the prethermal phenomenon. Since a larger $\alpha$ implies a better suppression of the low frequency components in the driving spectrum, the heating rate in panel (b) becomes smaller than panel (a), for a given fixed $\delta t_{\mathrm{max}}$. We note that it would be interesting to systematically investigate the relation between the heating rate and the hyperuniformity in the future.}

\end{document}